\documentclass[useAMS,usenatbib]{mn2e}

\newcommand{\etal}{\mbox{et\ al.\ }}

\usepackage{rotating}
\usepackage{url} 

\title[The discovery of GRB\,100316D]
{Discovery of the nearby long, soft GRB\,100316D with an associated supernova}

\author[Starling \etal 2010]
{R.L.C. Starling$^1$\thanks{rlcs1@star.le.ac.uk}, K. Wiersema$^1$, A.J. Levan$^2$, T. Sakamoto$^{3,4,5}$, D. Bersier$^6$, \and P. Goldoni$^{7,8}$, S.R. Oates$^9$, A. Rowlinson$^1$, S. Campana$^{10}$, J. Sollerman$^{11,12}$, \and N.R. Tanvir$^1$, D. Malesani$^{12}$, J.P.U. Fynbo$^{12}$, S. Covino$^{10}$, P. D'Avanzo$^{10}$, \and P.T. O'Brien$^1$, K.L. Page$^1$, J.P. Osborne$^1$, S.D. Vergani$^{13,14}$,
S. Barthelmy$^{15}$, \and D.N. Burrows$^{16}$, Z. Cano$^6$, P.A. Curran$^9$, M. De Pasquale$^9$, V. D'Elia$^{17,18}$, \and P.A. Evans$^1$, H. Flores$^{14}$, A.S. Fruchter$^{15}$, P. Garnavich$^{19}$, N. Gehrels$^{4}$, \and J. Gorosabel$^{20}$, J. Hjorth$^{12}$, S.T. Holland$^{4,21,3}$, A.J. van der Horst$^{22,23}$, \and C.P. Hurkett$^1$, P. Jakobsson$^{24}$, A.P. Kamble$^{25}$, C. Kouveliotou$^{22}$, N.P.M. Kuin$^9$, \and L. Kaper$^{25,26}$, P.A. Mazzali$^{27,28,29}$, P.E. Nugent$^{30}$, E. Pian$^{31,28}$, \and M. Stamatikos$^{4,32,33}$, C.C. Th\"one$^{10}$ and S.E. Woosley$^{34}$\\
$^1$Dept. of Physics and Astronomy, University of Leicester, University Road, Leicester LE1
7RH, UK.\\
$^2$Department of Physics, University of Warwick, Coventry CV4 7AL, UK.\\
$^3$Center for Research and Exploration in Space Science and 
Technology (CRESST).\\
$^4$NASA Goddard Space Flight Center, Greenbelt, MD 20771, USA.\\
$^5$Joint Center for Astrophysics, University of Maryland, 
Baltimore County, 1000 Hilltop Circle, Baltimore, MD 21250, USA.\\
$^6$Astrophysics Research Institute, Liverpool John Moores 
University, Birkenhead CH41 1LD, UK.\\
$^7$Laboratoire Astroparticule et Cosmologie, 10 rue A. Domon 
et L. Duquet, F-75205 Paris Cedex 13, France.\\
$^8$DSM/IRFU/Service D’Astrophysique, CEA-Saclay, 91191 Gif-sur-Yvette, France.\\
$^9$Mullard Space Science Laboratory, University College
London, Holmbury St. Mary, Dorking, Surrey RH5 6NT, UK.\\
$^{10}$INAF-Osservatorio Astronomico di Brera, Via Bianchi 46, 
I-23807 Merate (LC), Italy.\\
$^{11}$The Oskar Klein Centre, Department of Astronomy, Albanova, 
Stockholm University, SE-106 91 Stockholm, Sweden.\\
$^{12}$Dark Cosmology Centre, Niels Bohr Institute, Copenhagen
University, Juliane Maries Vej 30, 2100 Copenhagen \O, Denmark.\\
$^{13}$University Paris 7, APC, UMR7164 CNRS, 10 rue Alice Domon et 
L\'eonie Duquet, 75205 Paris Cedex 13, France.\\
$^{14}$GEPI, Observatoire de Paris, CNRS-UMR 8111, 5 place Jules 
Janssen, 92195 Meudon, France.\\
$^{15}$Space Telescope Science Institute, 3700 San Martin Drive, Baltimore, MD 21218 USA.\\
$^{16}$Department of Astronomy and Astrophysics, Pennsylvania State University, 525 Davey Lab, University Park, PA 16802, USA.\\
$^{17}$INAF-Osservatorio Astronomico di Roma, Via Frascati 33, I-00040 
Monteporzio Catone, Italy.\\
$^{18}$ASI-Science Data Centre, Via Galileo Galilei, 
I-00044 Frascati, Italy.\\
$^{19}$Physics Dept., University of Notre Dame, Notre Dame, IN 46556, USA.\\
$^{20}$Instituto de Astrof\'isica de Andaluc\'ia (IAA-CSIC), Glorieta 
de la Astronom\'ia s/n, E-18.008 Granada, Spain.\\
$^{21}$Universities Space Research Association, 10211 Wincopin Circle, Suite 500, Columbia, MD 21044-3432, USA.\\
$^{22}$NASA/Marshall Space Flight Center, Huntsville, AL 35812, USA.\\
$^{23}$NASA Postdoctoral Program Fellow.\\
$^{24}$Centre for Astrophysics and Cosmology, Science Institute, 
University of Iceland, Dunhagi 5, IS-107 Reykjavik, Iceland.\\
$^{25}$Sterrenkundig Instituut Anton Pannekoek, University of Amsterdam, P.O. number: 94249, 1090 GE Amsterdam, the Netherlands.\\
$^{26}$Laser Centre, VU University, De Boelelaan 1081, 1081 HV Amsterdam, the Netherlands.\\
$^{27}$Max-Planck-Institut f{\" u}r Astrophysik, 
Karl-Schwarzschild- Strasse 1, 85741 Garching, Germany.\\
$^{28}$Scuola Normale Superiore, Piazza Cavalieri 7, 56127 Pisa, Italy.\\
$^{29}$INAF – Oss. Astron. Padova, vicolo dell’Osservatorio 5, 35122 Padova, Italy.\\
$^{30}$Computational Cosmology Center, Lawrence Berkeley National 
Laboratory, 1 Cyclotron Road, Berkeley, CA 94720, USA.\\
$^{31}$Osservatorio Astronomico Di Trieste, Via G.B.Tiepolo, I - 34143 Trieste, Italy.\\
$^{32}$Center for Cosmology and Astro-Particle Physics (CCAPP) Fellow.\\
$^{33}$Department of Physics, The Ohio State University, 191 West Woodruff Avenue, Columbus, OH 43210, USA.\\
$^{34}$UCO/Lick Observatory, Department of Astronomy \& Astrophysics, University of California, Santa Cruz, CA95064, USA.
}

\begin{document}
\date{Accepted . Received ; in original form }

\pagerange{\pageref{firstpage}--\pageref{lastpage}} \pubyear{2010}

\maketitle

\label{firstpage}

\clearpage


\begin{abstract}
We report the {\it Swift} discovery of the nearby long, soft gamma-ray burst GRB\,100316D, and the subsequent unveiling of its low redshift host galaxy and associated supernova. We derive the redshift of the event to be $z = 0.0591 \pm 0.0001$ and provide accurate astrometry for the GRB-SN. We study the extremely unusual prompt emission with time-resolved $\gamma$-ray to X-ray spectroscopy, and find that the spectrum is best modelled with a thermal component in addition to a synchrotron emission component with a low peak energy. The X-ray light curve has a remarkably shallow decay out to at least 800~s. The host is a bright, blue galaxy with a highly disturbed morphology and we use Gemini South, VLT and HST observations to measure some of the basic host galaxy properties. We compare and contrast the X-ray emission and host galaxy of GRB\,100316D to a subsample of GRB-SNe. GRB\,100316D is unlike the majority of GRB-SNe in its X-ray evolution, but resembles rather GRB\,060218, and we find that these two events have remarkably similar high energy prompt emission properties. Comparison of the host galaxies of GRB-SNe demonstrates, however, that there is a great diversity in the environments in which GRB-SNe can be found. GRB\,100316D is an important addition to the currently sparse sample of spectroscopically confirmed GRB-SNe, from which a better understanding of long GRB progenitors and the GRB--SN connection can be gleaned.
\end{abstract}

\begin{keywords}
gamma-ray burst: individual: GRB\,100316D; supernovae: individual: SN\,2010bh
\end{keywords}

\section{Introduction}
\label{sec:intro}
The connection between long-duration gamma-ray bursts (GRBs) and Type Ic core-collapse supernovae (SNe) has long been established, beginning with the association of GRB\,980425 with SN\,1998bw \citep{Galama,Pian98}. Subsequent associations between nearby GRBs and spectroscopically-confirmed SNe include GRB\,030329/SN\,2003dh \citep{Hjorth,Stanek}, GRB\,031203/SN\,2003lw \citep{Malesani} and most recently GRB\,060218/SN\,2006aj \citep{Campana,Pian2}. These are spectroscopically confirmed examples of nearby GRB-SN associations (out to $z = 0.17$). The characteristic `bump' of a supernova has been noted in the photometric data for a dozen or more GRBs out to $z\sim1$ \citep[e.g.][]{Zeh,Ferrero,Dellavalle,Woosley,Tanvir}, while the majority of GRBs lie at higher redshifts where such signatures would be impossible to detect \citep[$\langle z \rangle = 2.2$][]{Jakobsson,Fynbo2}. The GRBs with SNe are therefore rare, but provide a crucial insight into the GRB--SN connection and the progenitors of long GRBs.

The GRBs with spectroscopically confirmed SNe are generally underluminous and subenergetic in comparison to a typical long GRB, though GRB\,030329 is a notable exception to this. The prompt emission of these bursts has a lower energy spectral peak than a typical GRB \citep{Kaneko} and they are suggested to have less relativistic outflows, or be viewed more off-axis. It has been suggested that the observed nearby ($z<0.1$), underluminous/subenergetic GRBs, including GRB\,100316D, may be drawn from a different population to the cosmological GRB sample, motivated by estimates of the local GRB rate several times greater than the rate of cosmological GRBs \citep[e.g.][]{Cobb,Chapman,Liang}. Radio observations of GRBs suggest that mildly relativistic ejecta and often high luminosities are the properties that distinguish the supernovae associated with GRBs from the non-relativistic ordinary Type Ib/c supernovae \citep{Kulkarniradio,Soderbergradio}.

GRB\,060218 was a landmark long-duration, low-luminosity, soft spectral peak event, detected by the {\it Swift} satellite \citep{Gehrels} with unprecedented multiwavelength coverage of the prompt emission. The prompt spectrum was found to comprise both the non-thermal synchrotron emission ascribed to most GRBs and thought to originate in the collision of fast-moving shells within the GRB jet \citep[e.g.][]{Rees}, and a thermal component. The presence of this thermal component, together with its evolution, led to the suggestion that we were observing the shock breakout of the supernova for the first time \citep{Campana,Waxman}. However, the non-thermal emission did not differ greatly from that of the X-ray flash class of soft GRBs and an outflow speed close to the speed of light could be inferred \citep{Toma}, alternatively suggesting this to be an extension of the typical GRB population and not requiring significantly slower ejecta or any special (off-axis) geometry \citep{Ghisellini}. \cite{Toma} speculate that the long-duration, low-luminosity events such as this one may point to a different central engine for these events compared with more typical GRBs: a neutron star rather than a black hole \citep[see also][]{Mazzali,Fan}. 

Also of relevance to this discussion are the supernovae for which no GRB was detected, yet for which mildly relativistic ejecta provide a good explanation for the radio data: SN\,2007gr \citep{Paragi} and SN\,2009bb \citep{Soderberg}. As GRB jets are thought to be highly collimated, it is expected that many will go undetected if the emission is directed away from our line of sight \citep[e.g.][]{Mazzali1}, and a fraction will also lie below current detector sensitivity limits. \cite{Soderberg} estimate the fraction of Type Ib/c SNe with central engines at about one per cent, consistent with the inferred rate of nearby GRBs.
In contrast, there are nearby GRBs for which an accompanying supernova could be expected but none has been detected to extremely deep limits. GRBs 060505 and 060614 \citep{Fynbo,Dellavalle,Gal-Yam} are the two best examples of this, highlighting the need for a greater understanding of the relationship between GRBs and SNe.

We present the discovery of a new GRB-SN, GRB\,100316D \citep{Stamatikos} associated with SN\,2010bh \citep{Wiersema2,Bufano,Chornock}. This is an unusually long-duration, soft-spectrum GRB positioned on a nearby host galaxy. Many of the properties of this GRB appear unlike those of typical GRBs, resembling rather GRB\,060218. In this paper we report accurate astrometry for the GRB-SN and the redshift of the underlying host galaxy. We examine the GRB prompt and afterglow emission observed with all 3 instruments on-board {\it Swift}, and the broad characteristics of the host galaxy as observed with the Gemini South telescope, the Very Large Telescope (VLT) and the Hubble Space Telescope (HST). We compare these properties with the GRB sample as a whole and with a subset of GRB-SNe, to understand the origins of the GRB emission and further our knowledge of the GRB--SN connection.

\section{Observations and analyses} 
\label{sec:obs}
\subsection{{\it Swift}} 
\label{sec:swiftobs}
On 2010 March 16 at 12:44:50 UT (hereafter T$_{\rm 0}$), the {\it Swift} Burst Alert Telescope (BAT, \citealt{Barthelmy}) triggered on and
slewed immediately to GRB\,100316D \citep{Stamatikos}. This was an image trigger, since no rapid, substantial rise in the count rate occurred,
but a possible low-level peak about 100~s long was noted in \cite{Stamatikos}.
The X-ray Telescope (XRT, \citealt{Burrows}) on-board {\it Swift} began observing the field at 12:47:08.1 UT, 137.7~s after
the BAT trigger, beginning with 10~s of Windowed Timing (WT) mode settling data and followed by 593~s of WT pointing data. After a time gap due to observing constraints, data-taking recommenced at T$_{\rm 0}$+33~ks with XRT in Photon Counting (PC) mode. The best X-ray position for this source is the astrometrically corrected XRT position of RA, Dec (J2000) = 07h 10m 30.63s, -56d 15' 19.7$''$, with an uncertainty of 3.7$''$ \citep[radius, at 90\% confidence][]{StarlingXRTpos}. {\it Swift's} UltraViolet-Optical Telescope (UVOT, \citealt{Roming}) began observing with the white filter 148~s after the BAT trigger and continued observations in u band, thereafter cycling repeatedly through all its lenticular filters.

The {\it Swift} data were processed and analysed with the {\it Swift} tools version 3.5, released as part of {\sc HEASOFT} 6.8, and the latest calibration data at 2010 March were used.

XRT WT spectra were grouped with a minimum of 20 counts in each bin, allowing the use of $\chi^2$ statistics, while for the PC spectrum 1 count per bin was required and Cash statistics used. We fit the X-ray spectra over the energy range 0.3--10 keV and the $\gamma$-ray spectra over the range 15--50 keV (as this is a soft event very few counts are present above 50 keV) using the spectral fitting package XSPEC \citep{Arnaud}. We used the X-ray absorption models {\sc phabs} and {\sc zphabs} for which photoionization cross-sections are taken from \cite{Verner} and Solar abundances from \cite{Wilms}. The Galactic absorption in the X-ray band is fixed at 7$\times$10$^{20}$ cm$^{-2}$ throughout \citep{Kalberla}. We tested for any differences in the spectral fitting results from use of different grade selections or different energy ranges (excluding 0.3--0.5 keV and/or including 50--150 keV) and confirm that these have no effect.

Optical/UV photometry was performed with the UVOT tool {\sc uvotsource}. At the position of the SN reported by \cite{Levan}, background subtracted count rates were extracted using
a circular source aperture of 2.5$''$. Source-free background was taken from a region situated outside of the host galaxy. The count rate was
aperture corrected to 5$''$ in order to be compatible with the
UVOT calibration \citep{Poole}. With UVOT we detect the host galaxy, but no new point sources. The 3$\sigma$ upper limit on any optical transient emission was taken to
be the background subtracted count rate plus three times
the count rate error and was converted into magnitude using
the zero points provided in \cite{Poole}. At
this time, galaxy subtraction cannot be performed since there
are no UVOT pre- or post-explosion observations.
The host galaxy count rate was determined using an elliptical source region, which enclosed the whole galaxy, and a source-free
background region. Photometry of the host was performed on a sum of all data from $\sim1\times10^5$~s
after the trigger onwards for each filter. Since the ellipse is similar in area
to a 5$''$ circular aperture (semi-major and -minor axes of 7.5$''$ and
5.5$''$ respectively), we have taken the standard UVOT co-incidence
loss correction to be the best estimate of the co-incidence loss,
and the standard UVOT zero points have been used to convert the
count rate into magnitudes.

\subsection{Gemini, VLT and HST} 
\label{sec:groundbased}
\begin{figure*}
\begin{center}
\includegraphics[width=16cm, angle=0]{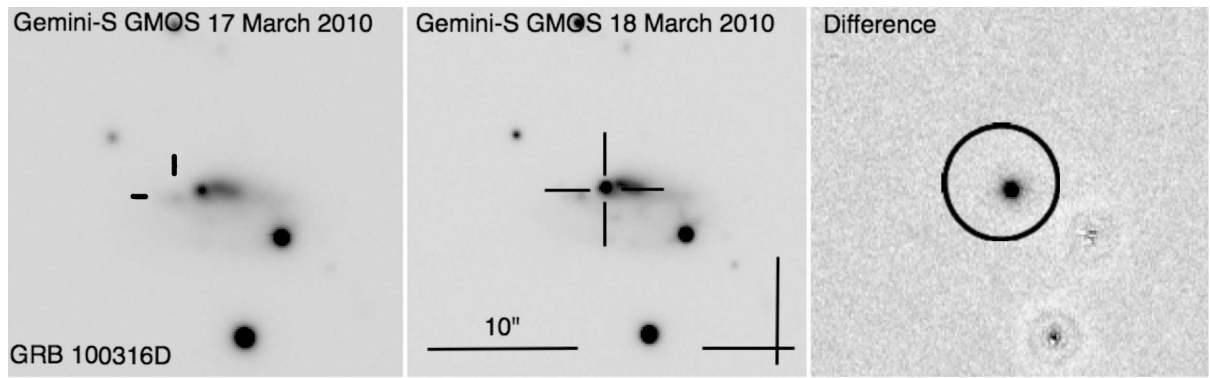}
\caption{Our Gemini-South discovery images of SN\,2010bh, associated with GRB\,100316D. The r-band images, obtained one night apart, clearly show the afterglow/SN superimposed on its host light. The GRB/SN position, image scale and North and East directions are indicated in the middle panel. The star forming region `A' is indicated in the left panel.
The result of PSF matched image subtraction is shown in the right hand panel, revealing
a brightening of the transient. The {\it Swift} XRT error circle is overlaid on this panel.}
\label{subtraction}
\end{center}
\end{figure*}

Our first optical observations of GRB\,100316D were obtained with Gemini-South, starting on March 16
23:53 UT. In these images we identified two point sources, consistent with the initial XRT position
\citep{Vergani}. We acquired spectroscopy of these with the 3-arm
echelle spectrograph X-shooter \citep{Dodorico}, mounted on Unit Telescope 2 of the VLT.
These sources were later shown to be compact H\,{\sc II} regions unrelated
to the supernova. In this paper we will make use of the spectra obtained
of
the brightest of these two regions, referred to as source `A' in
\cite{Vergani} (see Figure \ref{subtraction}). We obtained two spectra each of 1200~s exposure of source
A,
using a 5$''$ nod throw along the slit to achieve better sky
subtraction. We used slit widths of 1.0, 1.0 and 0.9$''$ for the UVB, VIS and NIR arms, respectively. Data were reduced using version 0.9.4 of the ESO X-shooter
pipeline \citep{Goldoni}. Calibration data (bias, dark, arc lamp, flatfield
and flexure control frames) were taken the same night. Extracted spectra
were flux calibrated using spectrophotometric standard star exposures, also
taken the same night, resulting in flux calibrated spectra spanning the ultraviolet
to the near-infrared ($\sim 300 - 2300$ nm) simultaneously.

We obtained further imaging of the localization with Gemini-South on the subsequent
three nights, using these we were able to identify a further
bright point source superimposed on the host galaxy \citep{Levan}. Image subtraction
performed between the different images showed a clear brightening in this source,
which we subsequently identified as the associated rising supernova \citep[][see Figure \ref{subtraction}]{Wiersema2}.

We obtained deep imaging of the position of GRB\,100316D using the FORS2
instrument on Unit Telescope 1 of the VLT. Seeing conditions during both
epochs were good, averaging 0.8$''$, but the second epoch observations
suffer from fairly bright background caused by high airmass
and a bright moon. The GRB-SN is clearly visible in both epochs.

Following the announcement of a SN discovery \citep{Wiersema2}, we triggered our
cycle~17 HST programme. Here we report the first epoch, obtained on March 25, $\sim$9 days after the BAT trigger.

The details of the ground-based and HST observations can be found in Table
\ref{table:opticallog}. The data were reduced using standard procedures in
{\sc IRAF}.

\begin{table*}
\caption{Log of Gemini-South, VLT and HST observations. \label{table:opticallog}}
\begin{tabular}{@{}rlcc@{}}
\hline
Date start  & Telescope/instrument + filter/arm & T$_{\rm exp}$ & Comments\\
(UT) & & (s) &\\
\hline
2010-03-16, 23:53 & Gemini-S/GMOS r & $5 \times 180$ & \\
\hline
2010-03-17, 00:58 & VLT/X-shooter UVB+VIS+NIR & $2 \times 1200$ & region A\\ \hline
2010-03-17, 23:55 & Gemini-S/GMOS r & $5 \times 180$ & \\
2010-03-18, 23:48 & Gemini-S/GMOS r & $5 \times 180$ & \\
2010-03-20, 00:28 & Gemini-S/GMOS griz & $3 \times 60$ & \\
\hline
2010-03-24, 00:07 & VLT/FORS2 R$_{\rm special}$ & 180 & \\
 & VLT/FORS2 I$_{\rm Bessel}$ & 180 & \\
 & VLT/FORS2 V$_{\rm high}$& 180 & \\ \hline
2010-03-25, 04:38 & HST/WFC3/UVIS F555W & $2\times 60$  & \\
\hline
2010-03-29, 01:11 & VLT/FORS2 V$_{\rm high}$& $6 \times 150$ & SN saturated\\
 & VLT/FORS2 R$_{\rm special}$& $17 \times 50$ & \\
 & VLT/FORS2 I$_{\rm Bessel}$& $9 \times 75$ & \\
 & VLT/FORS2 V$_{\rm high}$& $17 \times 50$ & \\
\hline
\end{tabular}
\end{table*}

\section{Astrometry and redshift}
Subtraction of early from late images taken with Gemini-South reveals a variable source inside the X-ray error circle which we conclude is GRB\,100316D/SN\,2010bh (Figure \ref{subtraction}). We find an accurate position for this variable source by astrometrically calibrating the images, using 190 USNO-B stars in the field. We determine the position of the residual
in the image subtraction by using a Gaussian centroiding. This gives a
position for the supernova of RA, Dec (J2000) = 07h 10m 30.558s, $-$56d 15' 20.18$''$
($\pm$ 0.26$''$). The error is dominated by the scatter of the USNO positions around the best
fitting polynomial of the astrometric calibration.
We refine this using an image taken with HST, to our best position of RA, Dec (J2000) = 07h 10m 30.530s, $-$56d 15' 19.78$''$ ($\pm$ 0.05$''$).

We detect a large number of nebular emission lines commonly found in
star forming regions with good signal to noise in the VLT X-shooter spectrum of source A (see Section \ref{sec:host}, Figure \ref{metallicity}). From the 12 brightest lines we measure the redshift $z = 0.0591 \pm 0.0001$.  Using the cosmology $\Omega_{\rm M} = 0.27$, $\Omega_{\Lambda} = 0.73$ and $H_{\rm 0} = 71$~km~s$^{-1}$~Mpc$^{-1}$, the luminosity distance is 261~Mpc.

\section{Temporal characteristics of the prompt emission} 
\label{sec:lightcurves}
\begin{figure}
\begin{center}
\includegraphics[width=7cm, angle=0]{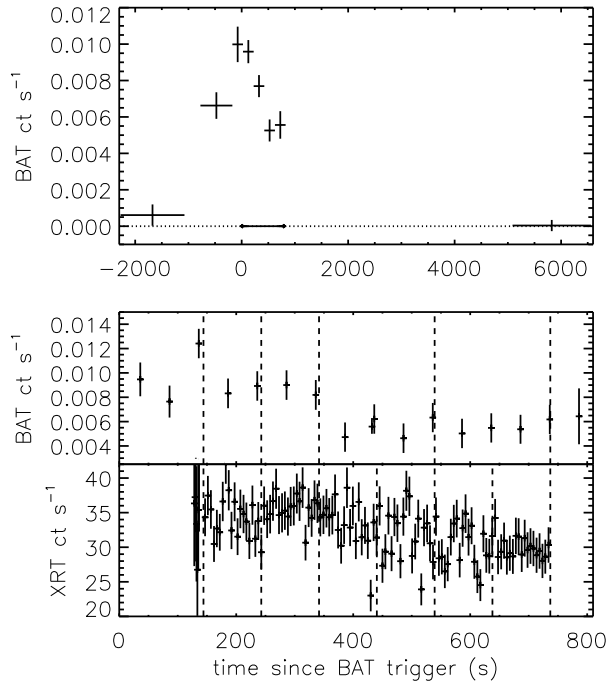}
\caption{Top panel: 14--195~keV long-term BAT light curve, comprising both event and survey binned data. The zero count-rate level is indicated with a dashed line, and the time range covered by the panels below is overlaid as a double-sided arrow. Lower two panels: zoom in on the 14--195~keV BAT event light curve and 0.3--10~keV XRT light curve from T$_{\rm 0}$ to T$_{\rm 0}$+810~s. The dashed lines indicate time slices adopted for spectral fitting.}
\label{BATlc}
\end{center}
\end{figure}
The BAT and XRT early emission light curves are shown in Figure \ref{BATlc}. The $\gamma$-ray emission was detected at the 10$\sigma$ level at T$_{\rm 0}$$-$475~s, prior to which there was no detected emission at 5$\sigma$ or above. The emission peaked at T$_{\rm 0}$$-$100~s, and decayed with an exponential decay constant of 750~s. Emission continued through the slew at T$_{\rm 0}$+750~s. BAT re-observed at T$_{\rm 0}$+5050~s from which time $\gamma$-ray emission was no longer detected.

The prompt emission duration of this GRB is one of the longest ever measured. GRB\,100316D was detected with BAT from $\sim$T$_{\rm 0}$$-$500~s to at least $\sim$T$_{\rm 0}$+800~s, hence the lower limit on the duration of GRB\,100316D is $\sim$1300~s. 
The fluence, derived from the 15--150~keV spectrum accumulated between T$_{\rm 0}$$-$475.0 and T$_{\rm 0}$+795.0~s since trigger, is $S_{\nu}$~=~(5.10 $\pm$ 0.39) $\times$10$^{-6}$ erg cm$^{-2}$. We note that as $\gamma$-ray emission continued after this time but {\it Swift} was unable to observe, this should be considered a lower limit to the burst fluence. The isotropic equivalent energy that follows from this is $E_{\rm iso} \ge$~(3.9$\pm$0.3)~$\times$10$^{49}$ erg. This is valid for the energy range 15--150 keV, but if we extrapolate down to 1 keV given that this is a soft burst, we find that the fluence and therefore $E_{\rm iso}$ may be higher by 50\% leading to $E_{\rm iso} \ge$~(5.9$\pm$0.5)~$\times$10$^{49}$ erg. We note that little flux is observed above 50 keV; the contribution to the isotropic energy from very high energy $\gamma$-rays is likely to be small. 

The BAT and XRT simultaneous coverage spans 603~s. The XRT count rate light curve decays very slowly with $\alpha = 0.13\pm0.03$ throughout this interval.

No variable optical or UV source has been detected during {\it Swift} observations with the UVOT, hampered by the bright underlying host galaxy \citep{Oates}. The light curves for each filter are consistent with a constant value, and we confirm that no variable source is detected with image subtraction using the full UVOT dataset to date. Upper limits on the magnitudes of any source at the GRB position are given in Table \ref{tab:uvotmags}.

\begin{table}
\caption{
UVOT 3$\sigma$ upper limits on the magnitude at the
position of the SN, from event and image data. We note that these are background
subtracted but not host-subtracted. The time ranges were chosen
to overlap with the simultaneous BAT-XRT coverage.}
\begin{tabular}{llll}
Filter& T$_{\rm mid}$ & T$_{\rm exp}$ & Magnitude \\
& (s since BAT trigger) & (s) & \\\hline
white & 195 & 93 & $>$21.3 \\
& 270 & 54 & $>$21.0 \\
& 597 & 19&$>$20.3 \\
v & 647 & 19 & $>$18.4 \\
b & 650 & 22& $>$19.6 \\
u & 324& 35 & $>$19.3 \\
 & 440 & 194 & $>$20.6 \\
 & 634 & 36 & $>$19.4 \\
uvw1 & 696 & 19 & $>$18.2 \\
uvm2 & 671 & 19 & $>$17.6 \\  \hline
\end{tabular}
\label{tab:uvotmags}
\end{table}

\section{Spectral characteristics of the prompt emission}
\label{sec:spec}

\begin{figure}
\begin{center}
\includegraphics[width=6cm, angle=-90]{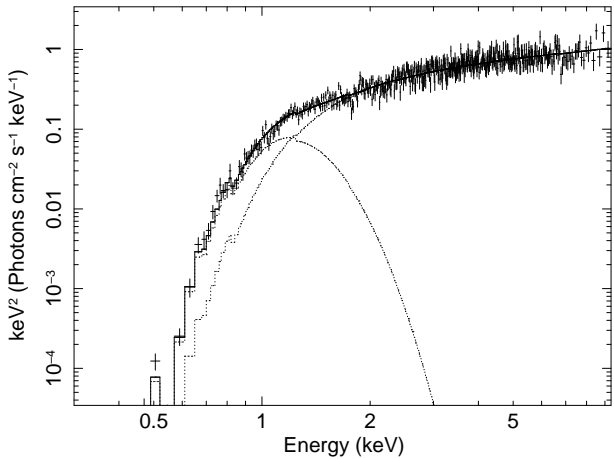}
\caption{Unfolded XRT WT time-averaged spectrum, $EF_E$ vs. $E$, with an absorbed power law plus blackbody model, showing the contributions of the two components.}
\label{XRTbbody}
\end{center}
\end{figure}

\begin{figure}
\begin{center}
\includegraphics[width=7cm, angle=0]{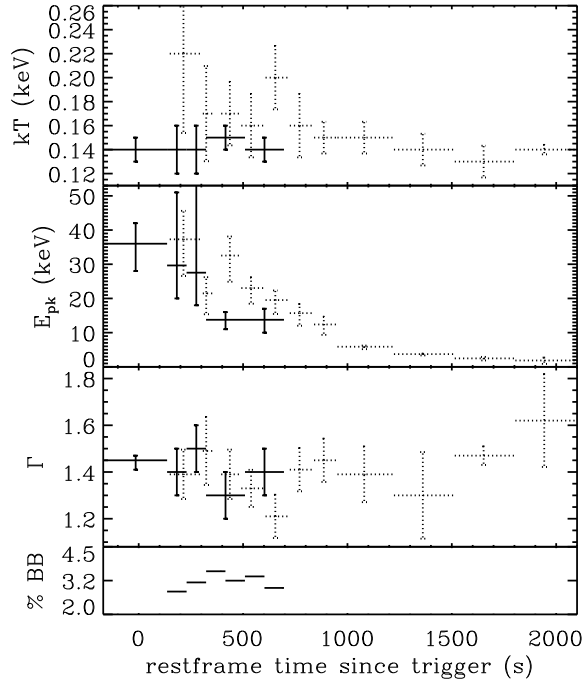}
\caption{Evolution of power law photon index $\Gamma$, peak energy $E_{\rm pk}$ and blackbody temperature $kT$ in the blackbody plus exponentially cut-off power law model fitted to the BAT-XRT spectra for GRB\,100316D (solid error bars), compared with the same model fitted to BAT-XRT data of GRB\,060218 (taken from Kaneko et al. 2007, dotted error bars) in the source rest-frames. Errors are 90\% confidence. The bottom panel shows the percentage contribution of the blackbody component to the total observed 0.3--10 keV X-ray flux, in time-sliced XRT spectral fits to GRB\,100316D.}
\label{Epkevolcf060218}
\end{center}
\end{figure}
We have carried out time-resolved spectroscopy during the 593~s of BAT-XRT overlap from T$_{\rm 0}$+144~s to T$_{\rm 0}$+737~s. For the X-ray data, we take six consecutive time slices, each of $\sim$99~s duration. We fit an absorbed power law model, and then include a further blackbody component (Figure \ref{XRTbbody}). In either case the X-ray spectral shape appears stable throughout the WT observations. The intrinsic absorption measured in both sets of fits is consistent at each epoch. We then fit the time-averaged WT spectrum consisting of data from T$_{\rm 0}$+144 to T$_{\rm 0}$+737~s since trigger to obtain the best constraints on the intrinsic column which we use in the following BAT-XRT joint fits. The time-averaged WT spectrum and the time-sliced WT spectra are adequately fit with an absorbed power law model, however the inclusion of a blackbody provides a significant improvement in the fits (F-test probabilities of 10$^{-10}$ and 10$^{-6}$ respectively),
requiring a larger intrinsic absorption column to accommodate a $kT = 0.14\pm0.01$ keV blackbody component while the power law photon index remains approximately the same. The blackbody component contributes $\sim$3\% of the total 0.3--10 keV X-ray flux in this model, and has a luminosity of (3--4)$\times$10$^{46}$ erg s$^{-1}$ corresponding to an emitting radius of 8$\times$10$^{11}$~cm. The blackbody luminosity and temperature (Figure \ref{Epkevolcf060218}) are consistent with remaining constant until at least T$_{\rm 0}$+800~s. 

In order to more rigorously test the significance of the blackbody component, we simulated 10000 spectra based on the absorbed power law model that best fits the time-averaged WT spectrum of GRB\,100316D (Table \ref{tab:specfits}). We fitted the simulated spectra with absorbed power law plus blackbody models to recover the chance probability that adding a blackbody component significantly improves the fit, following the methodology described in \cite{Hurkett}. There were no simulated spectra for which $\Delta\chi^2$ indicated an improvement in the fit comparable to that found in our data, implying that the blackbody component in the time-averaged WT spectrum of GRB\,100316D is $>4\sigma$ significant. This test gives us a high degree of confidence in our identification of a thermal component in the X-ray spectrum at early times. 

The joint BAT-XRT fits were performed with two 99~s and two 198~s consecutive time slices. We fit these spectra with an exponentially  cut-off power law model\footnote{defined as $A(E) = E^{-\Gamma} e^{-E(2-\Gamma)/E_{\rm pk}}$} with and without a blackbody component. A constant normalisation ratio between the BAT and XRT spectra (to account for calibration uncertainties in the response matrices of the different instruments) is not included in the reported fits since we found this to be 0.9$\pm$0.2, i.e. consistent with unity. The exponentially cut-off power law model provides a good fit with a stable low energy power law slope of $\Gamma \sim1.4$. The peak energy is at $\sim$30 keV in the first time interval and moves to lower energies with time (Figure \ref{Epkevolcf060218}).
To estimate the spectral shape during and prior to the BAT trigger we fitted the BAT spectrum from T$_{\rm 0}$$-$175 to T$_{\rm 0}$+144~s simultaneously with the time-averaged WT spectrum, given that no significant X-ray spectral evolution is observed in the interval 144--737~s. We stress that because the data are not simultaneous this fit provides only an indication of the spectral shape at early times. 
Spectral fitting results are given in Table \ref{tab:specfits}.

The latest time slice taken for our joint BAT-XRT fits also has full coverage in the u band with UVOT. We have included the u band 3$\sigma$ upper limit and refit the data with an absorbed and reddened blackbody+power law model. We obtain an acceptable fit, when the underlying spectrum is fixed in shape as listed in Table \ref{tab:specfits}, with an intrinsic extinction of $E(B-V) \sim 0.9$ mag or $A_v \sim 2.6$ mag using the Small Magellanic Cloud extinction law \citep[][fixing the Galactic extinction to $E(B-V) = 0.12$ mag according to \citealt{Schlegel}]{Pei}. This provides a lower limit on the host galaxy extinction along the line-of-sight to the GRB, only under the assumption that any optical transient emission lies on an extrapolation of the high energy spectrum. A measurement of the optical extinction along the line-of-sight to the GRB can be made with spectroscopy of the supernova, for example through the strength of interstellar medium Na I lines (Bufano et al. in preparation).
\begin{table*}
\caption{Spectral fits to the {\it Swift} BAT and XRT data with an absorbed power law (PL) or cut-off power law (CPL) model, with and without a blackbody (BB) component. Errors are quoted at the 90\% confidence level. Galactic absorption is fixed at 7$\times$10$^{20}$ cm$^{-2}$. \newline $^{t}$ These parameters are tied together for all time intervals.\newline $^*$ XRT data are not available during this time. We perform this fit only to provide an indication of the cut-off power law parameters over this time interval, using the time-averaged WT spectrum given that no significant X-ray spectral evolution is detected in the interval 144--736~s. Therefore this fit was performed separately rather than simultaneously with the 144--736~s interval BAT-XRT time-sliced fits.}
\begin{tabular}{l|lll |c| llll}
 & \multicolumn{3}{c}{PL or CPL}& & \multicolumn{4}{c}{PL$+$BB or CPL$+$BB}\\ \hline
T  & $N_{\rm H,int}$ & $\Gamma$ & $E_{\rm pk}$ & & $N_{\rm H,int}$ & $kT$& $\Gamma$ & $E_{\rm pk}$ \\ 
(s since BAT trigger) & (10$^{22}$ cm$^{-2}$)& & (keV)& & (10$^{22}$ cm$^{-2}$)&(keV)& & (keV) \\ \hline
XRT WT time-averaged:&\multicolumn{3}{c}{$\chi^2/\nu = 587.9/480$} & &\multicolumn{4}{c}{$\chi^2/\nu = 538.8/478$}\\ 
144-737&0.91$\pm$0.05 & 1.40$^{+0.04}_{-0.03}$&-&& 1.83$\pm$0.06& 0.14$\pm$0.01& 1.53$\pm$0.06&-\\ \hline
XRT WT time-sliced:&\multicolumn{3}{c}{$\chi^2/\nu = 830.9/754$}&&\multicolumn{4}{c}{$\chi^2/\nu = 782.1/742$}\\ 
144--243& 0.90$\pm$0.05$^{t}$ & 1.35$\pm$0.06& - &&1.8$\pm$0.2$^{t}$&0.14$\pm$0.02& 1.5$\pm$0.1&-\\
243--342& $''$ & 1.41$\pm$0.06& - &&$''$&0.14$\pm$0.02&1.6$\pm$0.1 &-\\
342--440&$''$ & 1.36$\pm$0.06& - &&$''$&0.15$\pm$0.02&1.5$\pm$0.1 &-\\
440--539&$''$ & 1.43$\pm$0.06& - &&$''$&0.14$^{+0.03}_{-0.02}$& 1.6$\pm$0.1&-\\
539--638&$''$ & 1.43$\pm$0.06& - &&$''$&0.13$\pm$0.02&1.5$\pm$0.1 &-\\
638--737&$''$ & 1.40$\pm$0.06& - &&$''$&0.14$^{+0.03}_{-0.02}$ &1.6$\pm$0.1 &-\\ \hline
XRT PC time-averaged:&\multicolumn{3}{c}{C-statistic/$\nu$ = 13.8/15}&&\multicolumn{4}{c}{C-statistic/$\nu$ = 15.4/15}\\
32906--508263 & 0.90 fixed & 3.5$\pm$1.0 & - && 1.83 fixed& - & 4.6$^{+1.6}_{-1.3}$& - \\ \hline
BAT:&\multicolumn{3}{c}{$\chi^2/\nu = 16.0/14$}&&\multicolumn{4}{c}{}\\
-175--144 & -& 2.1$\pm$0.4&-&& & & &  \\  \hline
BAT-XRT:&\multicolumn{3}{c}{$\chi^2/\nu = 617.4/518$}&&\multicolumn{4}{c}{$\chi^2/\nu =572.1/516$}\\
-175--144$^*$ &  0.90 fixed & 1.33$\pm$0.03 & 33$^{+7}_{-5}$ & &1.83 fixed & 0.14$\pm$0.01 & 1.45$^{+0.02}_{-0.04}$ & 34$^{+8}_{-6}$ \\
BAT-XRT:&\multicolumn{3}{c}{$\chi^2/\nu = 853.4/768$}&&\multicolumn{4}{c}{$\chi^2/\nu =806.1/760$}\\
144--243& 0.90 fixed& 1.25$\pm$0.08 & 25$^{+9}_{-5}$&&1.83 fixed&0.14$\pm$0.02& 1.4$\pm$0.1 & 28$^{+23}_{-8}$ \\
243--342& $''$& 1.32$\pm$0.08 & 22$^{+9}_{-5}$ && $''$&0.14$\pm$0.02 & 1.5$\pm$0.1 & 26$^{+36}_{-8}$\\
342--539& $''$& 1.24$^{+0.06}_{-0.07}$ & 14$^{+3}_{-2}$&&$''$&0.15$\pm$0.01 & 1.3$\pm$0.1 & 13$^{+3}_{-2}$ \\
539--737& $''$& 1.28$\pm$0.07 & 14$\pm$3 &&$''$& 0.14$\pm$0.01 & 1.4$\pm$0.1 & 13$^{+4}_{-3}$\\ \hline
\end{tabular}
\label{tab:specfits}
\end{table*}

\section{The late-time behaviour}
\label{sec:latetime}
The X-ray emission after 10$^{4}$~s decayed more steeply than did the emission during the first 800~s after the BAT trigger. 
Treating the early and late X-ray emission separately, we fit only the data after 10$^{4}$~s with a power law decay. 
This results in an acceptable fit, with $\alpha = 1.35 \pm 0.30$.

The XRT count rates for this source lie within the spread of the entire XRT GRB light curve sample at all observed times, however its shape is atypical. 
Four light curve types were identified in the {\it Swift} XRT catalogue, to which 76\% of their sample could be ascribed \citep{Evans}. 
GRB\,100316D appears not to fit any of these light curve types, if all the early data are included.

The X-ray spectrum from T$_{\rm 0}$+3$\times$10$^{4}$ to 5$\times$10$^{5}$~s allows only poor constraints on the parameters of an absorbed power law model, but we find it is significantly spectrally softer at this time than in the earlier WT data. Using Cash statistics, and fixing the intrinsic absorption to that found in the combined WT spectrum, we find the power law has a photon index of $\Gamma = 3.5 \pm 1.0$ (Table \ref{tab:specfits}). This is a very high value, or a very soft spectrum, that although not unique among {\it Swift} bursts \citep[see e.g.][]{Evans} is difficult to accommodate within the standard fireball model generally used to explain GRB afterglow emission \citep[e.g.][]{Sari}, but may possibly be explained with inverse compton scattering (e.g. Waxman et al. 2007). The temporal decay rate, however, is rather typical of GRB afterglows. 

We now turn to the optical regime, to estimate the supernova contribution at the epochs of our VLT data. 
We performed point spread function (PSF) matched image subtraction on the
FORS2 images, using the {\sc ISIS2} code (Alard 2000). For the second epoch V$_{\rm high}$ data we use
the short exposure time images, as the frames with longer exposure time show the supernova to be somewhat
saturated. The subtraction shows a residual at the position of the supernova associated with this GRB
\citep{Wiersema2} in the V and I bands, whereas the subtraction is too noisy in the R band. This
residual shows that the supernova is fainter in the second FORS2 epoch than in the first epoch, at least in
V and I, suggesting that the supernova peaked in these bands between 8 and 13 days after burst. This is
similar to the SN\,2006aj (GRB\,060218), which peaked at $10.2 \pm 0.3$ days after burst in V \citep{Ferrero,Mirabal,Modjaz,Sollerman}.

\section{The host galaxy}
\label{sec:host}
The deep Gemini, VLT and HST observations show a bright, blue galaxy, covering
the 3.7$''$ radius refined XRT error circle (Figures \ref{subtraction} and \ref{GRBSNehosts}). The galaxy has a highly disturbed morphology, but is
dominated by a central bright region, possibly the galaxy nucleus, and an arc-like shape resembling a
spiral arm. The galaxy hosts a large number of knots, or blobs, likely giant H\,{\sc II} regions. Colour
differences between these knots can be discerned, caused by differences in age, metallicity, reddening
or a combination of these. The supernova itself is located on top of a bright, blue knot, close (in
projection) to the possible galaxy nucleus.
The UVOT also observed the host galaxy in all filters and approximate magnitudes, uncorrected for extinction and including an unknown level of contamination by the afterglow and/or SN, are reported in Table \ref{tab:uvothostmags}. 

It has long been contended
that the properties of the host galaxies of GRBs can provide additional clues to the progenitor evolution,
through determination of properties such as the age, metallicity and alpha element enrichment of the
stellar population. The low redshift GRB-SNe are of particular interest, as they allow one to obtain
spatially resolved spectroscopy \citep[e.g.][]{Christensen}. The derived properties can then be
meaningfully compared to those of `normal' core-collapse supernova hosts \citep{Fruchter,Svensson} or other star forming galaxy populations. 
For these reasons we analyse the X-shooter spectrum we obtained of a bright H\,{\sc II} region
within the host, which we call source A \citep{Vergani}, spatially distinct from the supernova region (Figures \ref{subtraction} and \ref{GRBSNehosts}). We choose to use this
spectrum rather than the X-shooter spectra of the supernova itself, as in
these spectra the contribution of the supernova to the spectrum is large and
can not be easily separated from the underlying H\,{\sc II} region lines.
Source A is located near the supernova (projected distance approximately 2\,kpc)
and we may expect the properties of this region to give a meaningful insight into the
properties of the host as a whole.

\begin{figure}
\begin{center}
\includegraphics[width=7cm, angle=0]{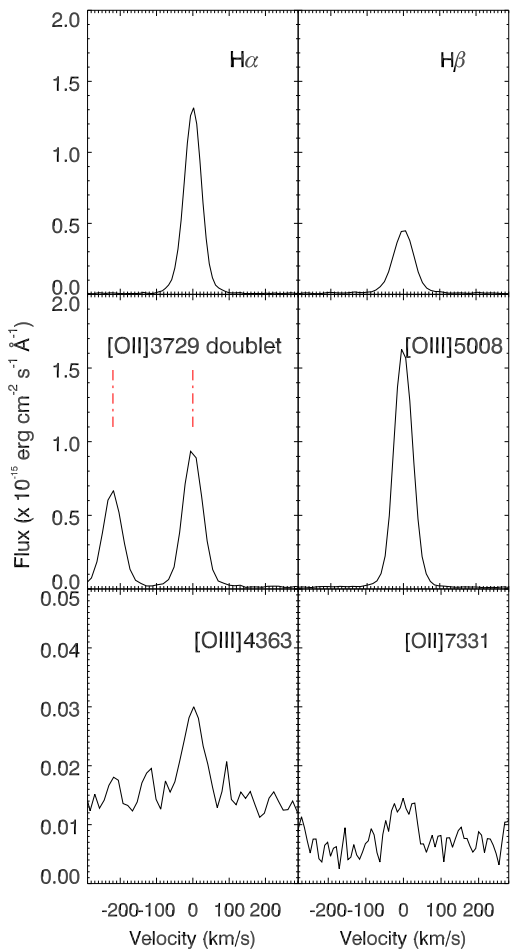}
\caption{A selection of emission lines which are most important for the oxygen
abundance analysis, detected in the X-shooter spectrum of the source `A' HII region in the host galaxy, $\sim$2kpc projection from the GRB/SN location. The top two panels show the Balmer H$\alpha$ and H$\beta$ lines used for the
determination of the reddening. The middle panels show the strong lines of the [O\,{\sc II}]~$\lambda$$\lambda$3727,3729
doublet and [O\,{\sc III}]~$\lambda$5008. The zero velocity point for [O\,{\sc II}]~$\lambda$$\lambda$3727,3729 is (arbitrarily) chosen to be the centre of the redmost component of the doublet.
The lower panel shows the temperature sensitive lines of [O\,{\sc III}] and [O\,{\sc II}] (of the latter only one member
of the doublet is shown) from which we derive $T_e ({\rm O}^{2+})$ and $T_e ({\rm O}^{+})$ respectively.}
\label{metallicity}
\end{center}
\end{figure}

We detect a large number of nebular emission lines commonly found in star forming regions; a selection of those used here (which are all unresolved) is shown in Figure \ref{metallicity}. We use the Balmer H$\alpha$ and H$\beta$ line fluxes to measure a
value for the Balmer decrement (jointly fitting the stellar Balmer absorption component and the nebular emission line component), assuming case B recombination
\citep[e.g.][]{Osterbrock,Izotov}. Assuming an electron temperature
of 10$^4$~K we compute the extinction (we find $E(B-V) = 0.178$ mag for the
combination of Galactic reddening of $E(B-V)_{\rm gal} = 0.07$ mag and intrinsic reddening in source A), and deredden the spectrum accordingly.

We compute the average electron density $n_e$ using the [S\,{\sc II}] doublet
and the [O\,{\sc II}] doublet: both of the doublet flux ratios are consistent with their
lower limits, so we assume a value of $n_e = 100$ cm$^{-3}$ \citep[see e.g.][]{Osterbrock}. The electron
temperature sensitive [O\,{\sc III}]~$\lambda$4363 and [O\,{\sc II}]~$\lambda$$\lambda$7320,7331
doublet are detected, from which we calculate the electron temperatures for
O$^{2+}$ and O$^{+}$ directly \citep[e.g.][]{Osterbrock}, at $11880 \pm 800$~K and
$10400 \pm 1100$~K, respectively (errors are 1$\sigma$). Ionisation corrections to account for the
presence of O$^{3+}$ were found to be negligible, also evidenced by the non-detection of He\,{\sc II} $\lambda$4686, so we determine the oxygen
abundance in region A as O/H~=~O$^{+}$/H$^{+}$~+~O$^{2+}$/H$^{+}$. Using the
measured line fluxes, densities and temperatures we arrive at an electron
temperature based oxygen abundance of 12~+~log(O/H)~$= 8.23 \pm 0.15$ in
region A (0.37 times Solar metallicity, assuming a solar oxygen abundance of 12 + log(O/H) = 8.66; Asplund et al.~2004). We note that a similar oxygen abundance has been derived from spectra at the SN site \citep[see][]{Chornock}.

In the spectrum we detect Balmer absorption lines underlying the nebular
emission. This is common in GRB host galaxy spectra, and can provide a useful age tracer, providing the spectra have    
sufficient resolution (e.g. Wiersema et al. in preparation). We fit the H$\delta$ absorption component, and find an approximate age for
the dominant stellar population of $\sim30$ Myr, assuming continuous star formation \citep{Gonzalez}. We also note the absence of obvious Wolf-Rayet star emission line features in source A. 
The detection of the bright H$\alpha$ line allows us to estimate the star formation rate at location A: the
H$\alpha$ line luminosity is a
tracer of recent star formation frequently used in nearby GRB host galaxies. We use the expression
$SFR_{{\rm H}\alpha} = 7.9 \times 10^{-42} L_{{\rm H}\alpha}$ \citep{Kennicutt}. Using the de-reddened flux
from the X-shooter spectrum we find $SFR_{{\rm H}\alpha} = 0.17$ M$_{\odot}$ yr$^{-1}$. Note that 
no aperture correction is performed, so this value is formally a lower limit.

A more detailed analysis of the host galaxy will appear in Flores et al. (in preparation).

\begin{table}
\caption{
UVOT observed magnitudes and 1$\sigma$ errors
for the host galaxy. Photometry is performed
as described in Section \ref{sec:swiftobs}. N.B. any afterglow or supernova contribution is uncertain
so could not be removed, however no variability is seen in
subtraction of late-time images from earlier epochs.
}
\begin{tabular}{lllll}
Filter& T$_{\rm start}$ & T$_{\rm stop}$& T$_{\rm exp}$ & Magnitude \\
& \multicolumn{2}{c}{(s since BAT trigger)} & (s) & \\\hline
 v &1.1$\times10^5$ &6.1$\times10^5$& 3229 & 17.71$\pm$0.03 \\
b & 1.1$\times10^5$& 6.1$\times10^5$&3578 & 18.29$\pm$0.03 \\
u &1.1$\times10^5$&6.1$\times10^5$ & 4325 & 18.02$\pm$0.03 \\
uvw1 & 1.1$\times10^5$&6.3$\times10^5$& 8851& 18.07$\pm$0.03 \\
uvm2 &1.1$\times10^5$ &6.1$\times10^5$& 8332 & 17.99$\pm$0.03 \\
uvw2 &1.1$\times10^5$ &6.1$\times10^5$& 16226& 18.00$\pm$0.02 \\ \hline
\end{tabular}
\label{tab:uvothostmags}
\end{table}
\begin{figure}
\begin{center}
\includegraphics[width=6.5cm, angle=-90]{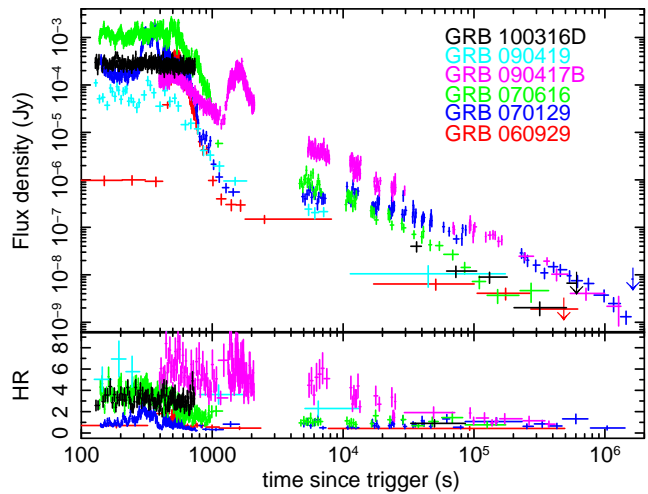}
\caption{Upper panel: X-ray light curves in flux density at 1.7 keV against time since trigger for a selection of {\it Swift} XRT-observed very long prompt emission duration ($>$400~s) GRBs (excluding 060218). 
Lower panel: (1.5--10 keV)/(0.3--1.5 keV) hardness ratios derived from the XRT count rate light curves.
}
\label{XRTlcs_long}
\end{center}
\end{figure}

\section{Discussion} 
\label{sec:discuss}

GRB\,100316D is an atypical gamma-ray burst both in its temporal and spectral behaviour. The very soft spectral peak and extended and slowly decaying flux emission are highly unusual among the prompt emission of GRBs \citep[e.g.][]{Sakamoto}. The estimated total isotropic energy ($E_{\rm iso} \ge$~(5.9$\pm$0.5)~$\times$10$^{49}$ erg), when considered with the 90\% confidence range of spectral peak values of 10--42~keV we measure in Section \ref{sec:spec}, places GRB\,100316D on the low $E_{\rm iso}$--low $E_{\rm pk}$ tail of the Amati correlation which relates $E_{\rm pk}$ and $E_{\rm iso}$ for the majority of typical GRBs \citep{Amati1}. The notable outlier to this relation is GRB\,980425 (SN\,1998bw) \citep[e.g.][]{Kaneko,Amati}.
We also compare this GRB with the relation reported by \cite{Gehrels2} between XRT flux at 11~h (here, (3.2$\pm$0.8)$\times$10$^{-13}$ erg cm$^{-2}$ s$^{-1}$) and BAT fluence and find, using the lower limit to the fluence, that it lies along the correlation with values that are typical for the sample of {\it Swift} GRBs reported in \cite{Evans}. 

GRB\,100316D is among the longest duration {\it Swift} GRBs ($T_{\rm 90}\ge1300$~s), after GRB\,090417B ($T_{\rm 90}\sim2130$~s) and GRB\,060218 ($T_{\rm 90}\sim2100$~s). GRBs with durations of 400~s or longer, as quantified using the $T_{\rm 90}$ parameter, account for only 1.4\% of the {\it Swift} sample observed to date. Owing to their long prompt emission durations, these GRBs are more likely to have exceptional broadband coverage, lending themselves to prompt emission studies. We show the X-ray light curve and hardness ratio of GRB\,100316D in relation to the $T_{\rm 90}>400$~s XRT-observed {\it Swift} events in Figure \ref{XRTlcs_long} (omitting GRB\,060218 as it can be seen in Figure \ref{XRTlcs_GRBSNe}; data are taken from the {\it Swift} XRT GRB light curve repository\footnote{\url{http://www.swift.ac.uk/xrt_curves/}}, \citealt{Evans1}).

This category comprises a number of different light curve types, from those which are long due to triggering on a precursor or the presence of a broad flare to the very gradually decaying prompt emission of GRB\,100316D. The {\it Swift} subsample shown here indicates two GRBs, in addition to GRB\,060218, whose emission resembles GRB\,100316D both in temporal decay and in spectral hardness: GRBs 070616 \citep{Starling070616} and 090419 \citep{Stratta}. We note
that GRB\,090417B is also spectrally hard at early times, but this may be
attributed to strong X-ray flaring (Holland et al. in preparation).

GRB\,070616 was particularly well sampled, with $\gamma$-ray through optical coverage at early times clearly showing the movement of the spectral peak energy of the Band function \citep{Band} and the occurrence of additional spectral softening \citep{Starling070616}. However, the spectrum was much harder than that of 100316D - observed by both {\it Swift} BAT and {\it Suzaku} WAM, and neither the redshift nor the presence of any SN could be established as the source lay close to a bright star.

We now discuss the properties of GRB\,100316D in the context of GRB-SNe, paying particular attention to GRB\,060218.

\subsection{Comparison to GRBs with accompanying supernovae: high energy emission}
\label{sec:cfGRBSNe}
\begin{figure*}
\begin{center}
\includegraphics[width=12cm, angle=-90]{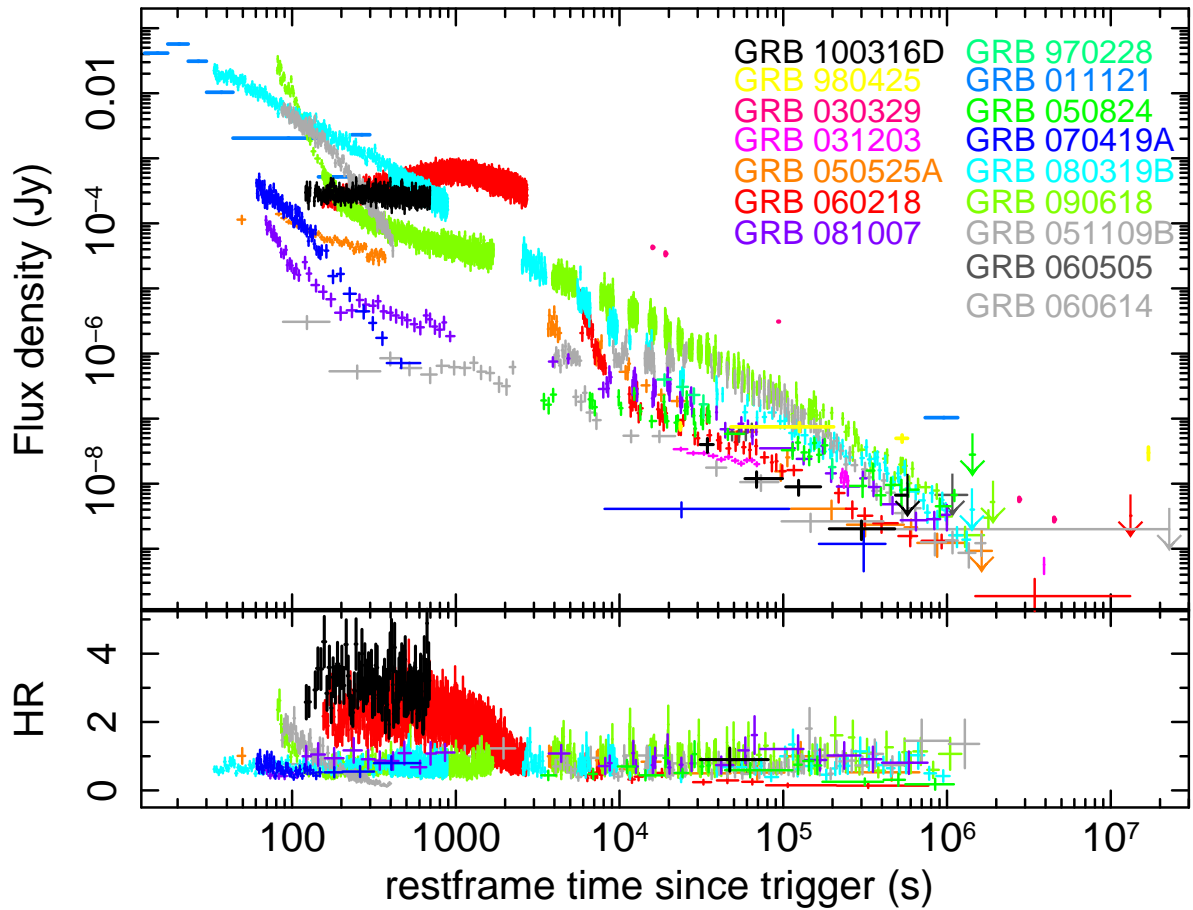}
\caption{X-ray light curves against time since trigger in flux density at 1.7 keV for a selection of long GRBs with associated 
supernovae (either spectroscopically confirmed - generally coloured in shades of red with GRB\,100316D in black, or photometrically determined - shades of blue and green), two nearby long GRBs with no observed supernovae to 
deep limits and a further long GRB likely at $z\le0.1$ with no reported supernova (grey shades). The lower panel shows the hardness ratios 
for the {\it Swift} XRT-observed GRBs, derived from the 1.5--10 keV and 0.3--1.5 keV count rate light curves in the same colour scheme.\newline
References for the data sources: 100316D$^2$, 980425 \citep{Pian}, 030329 \citep{Willingale}, 031203 \citep{Watson}, 050525A \citep[][and $^2$]{Blustin}, 060218$^2$, 081007$^2$, 970228 \citep{Costa,DePasquale}, 011121 \citep{Piro}, 050824$^2$, 070419A$^2$, 080319B$^2$, 090618$^2$, 051109B$^2$, 060505$^2$, 060614$^2$.}
\label{XRTlcs_GRBSNe}
\end{center}
\end{figure*}

In Figure \ref{XRTlcs_GRBSNe} we show the X-ray light curves and hardness ratios of a subset of the GRBs with associated SNe (either spectroscopically or photometrically confirmed). On this plot we also include GRB\,051109B which was tentatively associated with a very low redshift \citep[$z = 0.08$,][]{Perley} host galaxy and has no reported SN that we are aware of, and the well-studied nearby GRBs 060505 and 060614 for which SN searches to very deep limits registered no detections \citep{Fynbo,Dellavalle,Gal-Yam}. The prompt emission shows a broader range of observed fluxes than the late-time decays, perhaps suggesting they are unrelated or that an extra component is present in the early emission which contributes differently in each GRB. The majority of the GRB-SNe (and GRBs without SNe shown here) light curves decay steeply over the first 1000~s, with GRBs 100316D and 060218 being the only exceptions. The hardness ratios reveal that GRB\,100316D and GRB\,060218 have a similar spectral hardness evolution, clearly different from those of most other GRB-SNe which remain at a stable and softer spectral shape throughout the prompt and late-time emissions. GRBs 060614 (without a SN) and 090618 (photometrically discovered SN) also transition from hard to soft, but do this earlier, over the first 200~s during their steep decays. We note, however, that X-ray data during the early emission for a number of the GRB-SNe, including the spectroscopically confirmed GRB-SNe 980425, 030329, 031203 and 050525A, are either not available or inadequate for spectral studies.  

\subsection{Comparison to GRBs with accompanying supernovae: host galaxies}
\label{sec:cfGRBSNehosts}
\begin{figure*}
\begin{center}
\includegraphics[width=16cm, angle=0]{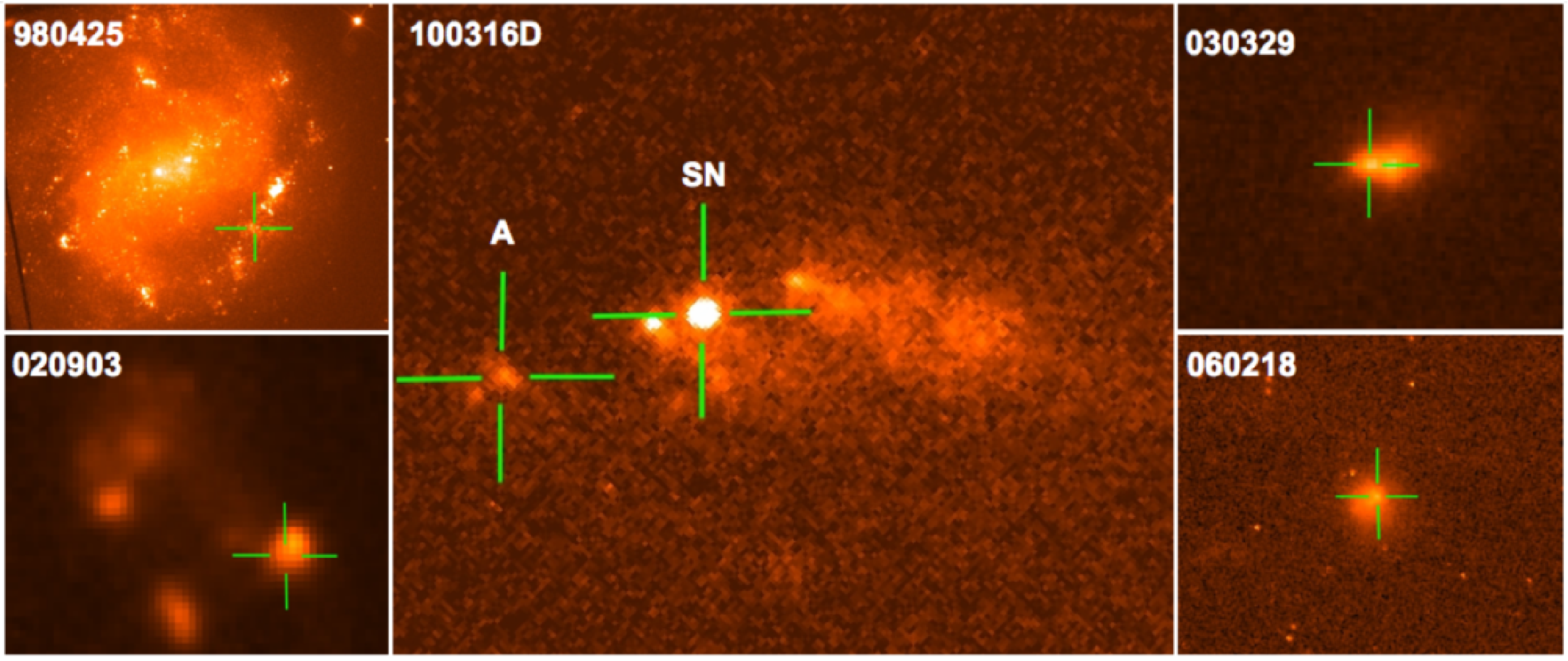}
\caption{A mosaic of the host galaxies of spectroscopically confirmed supernovae, associated with GRBs, and imaged by
the Hubble Space Telescope. The centre image
shows the host of GRB\,100316D, while other examples are shown in smaller insets. The physical scale
across each image is approximately 7~kpc, and the positions of source A and of the SN are marked with
crosshairs. Note, in the case of GRB\,060218 and GRB\,100316D there is still some contamination from the
supernova at the time of the image.}
\label{GRBSNehosts}
\end{center}
\end{figure*}

Nearby GRBs show a large variety in their host galaxy properties: the prototype GRB-SN, GRB\,980425 and SN\,1998bw, occured in a dwarf spiral, in a small H\,{\sc II} region right next to a very large star formation
complex; whereas GRB\,060218 and SN\,2006aj, occured in a very faint, blue compact
dwarf galaxy \citep{Wiersema1}. This diversity is illustrated in Figure \ref{GRBSNehosts}.

Assuming that the properties of source A are at least roughly representative of the host galaxy as a whole, we can compare the
spectroscopic properties with those of the other nearby GRB-SN hosts. Electron temperature based oxygen abundances have been 
derived for several of these, see Table \ref{table:grbsnhosts}. The case of GRB\,980425 is of particular interest: besides many
similarities between SN\,1998bw and SN\,2010bh there are also several similarities in host galaxy properties. The $T_e$          
metallicity measured at the location of the GRB and at the location of a nearby, bright, WR star region \citep{Hammer} are
clearly similar to the metallicity measured for the H\,{\sc II} region A in the host of GRB\,100316D. In addition, the size and       
brightness (and possibly morphology) of these two hosts are comparable, though an important difference between the host         
of 980425 and 100316D is evident from Figure \ref{GRBSNehosts}: the morphology of the host of 100316D is highly disturbed, possibly indicative of a        
recent merger.                                    

The host of GRB\,980425 has been studied using integral field spectroscopy \citep{Christensen}, showing that the H\,{\sc II} region in which the GRB exploded is similar in mass, SFR, reddening and line equivalent width to other H\,{\sc II} regions in this host, with the exception of a bright WR-star rich H\,{\sc II} region 800 pc away from the GRB site. In terms of metallicity, the WR region and GRB site have a somewhat lower metallicity than other H\,{\sc II} regions in this host \citep{Christensen}. If this situation is comparable to the host of GRB\,100316D, we may expect that the properties measured above of source A are indeed representative of most of the host galaxy properties. 

\begin{table*}
\caption{Host galaxies of nearby GRB-SNe. The absolute magnitudes $M_B$ were taken from the compilation of Levesque et al.~(2010).\label{table:grbsnhosts}}
\begin{tabular}{@{}lllll@{}}
GRB     & $T_e$ oxygen abundance             &  Host type      & Absolute magnitude & References\\
        & (12 + log(O/H))                    &                 & $M_B$              &           \\
\hline
980425  & 8.25 (GRB site)                    & Dwarf spiral    &  $-$17.6             &  Sollerman et al. (2005);         \\
        & 8.39 (nearby WR region)            &                 &                    &   Hammer et al. (2006)                             \\
020903  & 7.97                               & Irr             &  $-$18.8             & Hammer et al. (2006)          \\
030329  & 7.72                               & Irr             &  $-$16.5             &  Levesque et al. (2010)  \\
031203  & $8.02 \pm 0.15$ (integrated)       & Irr             &  $-$21.0             & Prochaska et al. (2004) \\
060218  & $7.54^{+0.16}_{-0.1}$  (integrated)& Irr             & $-$15.9              & Wiersema et al. (2007) \\
100316D & $8.23 \pm 0.15$ (source A)         & Spiral? Irr?      &   $-$18.8                 & This work \\
\hline
\end{tabular}
\end{table*}

\subsection{Comparison to GRB\,060218} \label{sec:cf060218}
In Section \ref{sec:cfGRBSNe} we found that among the GRB-SNe, there is a clear commonality in the X-ray behaviour of 100316D and 060218. Comparing the prompt emission spectra in Section \ref{sec:spec}, we again find parallels in that both bursts seem to require a similar thermal component in addition to the typical GRB synchrotron emission for which the synchrotron peak is observed to move to lower energies with time. The time-evolving prompt spectral parameters are compared in Figure \ref{Epkevolcf060218}.
Both events have low isotropic equivalent energies, of order 4$\times$10$^{49}$ erg, and very little flux above 50 keV. 

These two events lie in apparently quite different environments: 060218 in a faint, compact dwarf \citep{Wiersema1,Modjaz} and 100316D in a luminous, disturbed possibly spiral host galaxy. The metallicities are also different, with the host of 100316D likely being more metal-rich (though the metallicity is below Solar). A prompt optical component, as observed in 060218, could not be detected in 100316D, but conditions for its detection were far less optimal in this case due to the superposition on a brighter host galaxy and the possibility of a higher extinction along the line-of-sight. The higher intrinsic X-ray column density we measure for 100316D compared with similar modelling of 060218 does not necessarily imply a higher optical extinction \citep{Watson2,Campana2}. 

In summary, both 100316D and its predecessor 060218 are nearby ($z=0.059, 0.033$), long-duration ($T_{90} \ge$1300~s,~2100~s), initially relatively constant in X-ray flux (Figure \ref{XRTlcs_long}), spectrally soft (very few or no counts above 100 keV; low $E_{\rm pk}$, Figure \ref{Epkevolcf060218}), subenergetic (both have $E_{\rm iso}\sim 4\times 10^{49}$ erg) GRBs with an associated SN. These two events show similar prompt emission properties and stand out among the GRB-SNe subsample considered here for their unusual X-ray evolution. However, their host galaxies are rather different in morphology and metallicity, with the host of 100316D more closely resembling the host of 980425.

The thermal X-ray component, with a luminosity, temperature and radius similar to that observed in 060218, dictates that the shock break-out of the supernova must be considered. The optical/UV thermal component observed in 060218 is not seen here: if the extinction is similar we would expect a shock breakout similar to 060218 to be two or more times fainter in 100316D and with the relatively brighter host galaxy it is perhaps no surprise that early optical emission is not detected with UVOT. The presence of a thermal component in the prompt $\gamma$-ray emission of BATSE (Burst And Transient Source Experiment) GRBs, in addition to non-thermal emission, was proposed by \cite{Ryde}. The discovery of thermal components in the {\it Swift} XRT X-ray spectra shown here for 100316D and in \cite{Campana} for 060218 will therefore be important to consider in future studies of GRB prompt emission.
The dominant component of the high energy emission in 100316D remains the synchrotron-like non-thermal spectrum common to all types of GRB (with and without SNe) thought to originate in internal shocks in a relativistic jet. The long duration of the early X-ray emission is curious, and exceedingly rare, perhaps suggesting a greater reservoir of material is available to feed the central engine and prolong its activity.

The discovery of GRB\,100316D and its associated supernova SN\,2010bh and the analyses of its high energy emission and host galaxy properties presented here illustrates the diversity in GRB-SNe characteristics that must be understood if we are to fully appreciate the relationship between GRBs and core-collapse SNe.

\section{Acknowledgments}
This work made use of data supplied by the UK {\it Swift} Science Data Centre at the University of Leicester. Based on observations made with ESO Telescopes at the La Silla or Paranal Observatories under
programme IDs 084.D-0939, 083.A-0644 and 084.A-0260(B). Based on observations made with the NASA/ESA Hubble Space Telescope, obtained at the Space Telescope Science Institute, which is operated by the Association of Universities for
Research in Astronomy, Inc., under NASA contract NAS 5-26555: these observations are
associated with programme 11709. We thank STScI staff for their efforts in implementing the HST ToO observation,
particularly Alison Vick. We acknowledge the wider {\it Swift} team for
their many contributions. RLCS, KW, AR, JPO, KLP and PAE acknowledge financial support from STFC. Financial support of the British Council and Platform Beta Techniek through the Partnership Programme in Science (PPS WS 005) is gratefully acknowledged (KW). The Dark Cosmology Centre is funded by the Danish National Research Foundation. DNB acknowledges NASA contract NAS5-00136. AJvdH was supported by an appointment to the NASA Postdoctoral Program at the MSFC, administered by Oak Ridge Associated Universities through a contract with NASA.

\bsp

\label{lastpage}

\end{document}